\documentclass[
floatfix,
aps,prb,amsmath,
twocolumn,
superscriptaddress,
]{revtex4}

\usepackage{bm}

\usepackage{epsfig}
\usepackage{epsf}
\usepackage{array}

\def\e{\varepsilon}

\def\mls{\delta_1}
\def\cV{{J_{12}}}
\def\cJ{{\cal J}}

\begin{document}

\title{Transport spectroscopy of Kondo quantum dots coupled
  by RKKY interaction}

\author{Maxim G. Vavilov}\altaffiliation[Present address: ]
{Department of Applied Physics, Yale University, New Haven, CT06520}
\affiliation{Center for Materials Sciences and Engineering,
  MIT, Cambridge, MA 02139}
\author{Leonid I. Glazman} \affiliation{Theoretical Physics Institute,
 University of Minnesota,
  Minneapolis, MN 55455}
\date{May 30, 2005}
\begin{abstract}
  We develop the theory of conductance of a quantum dot which carries
  a spin and is coupled via RKKY interaction to another spin-carrying
  quantum dot. The found dependence of the differential conductance on bias and
  magnetic field at fixed RKKY interaction strength may allow one to
  distinguish between the possible ground states of the system.
  Transitions between the ground states are achieved by tuning the
  RKKY interaction, and the nature of these transitions can be
  extracted from the temperature dependence of the linear conductance.
  The feasibility of the corresponding measurements is evidenced by
  recent experiments by Craig {\it et al}.~\cite{Craig}
\end{abstract}
\maketitle
%\markright{%\hspace{12.65cm}
%\hfill\small\sl \today ~ -- ~\rm p.~}
%%%change%%%abstract

The exchange interaction between a localized electron and itinerant
electrons of a Fermi sea leads to the Kondo effect.~\cite{Hewson}
Recently, the Kondo effect was observed in the quantum dot setting, where it causes
an anomalously high conductance at low
temperatures.~\cite{Kondoreview} The itinerant electrons not only
screen an impurity spin, leading to the Kondo effect, but also give
rise to the Ruderman-Kittel-Kasuya-Yosida (RKKY)
interaction between localized spins.~\cite{Kittel} The interplay between the Kondo
screening and RKKY interaction remains at focus of investigation
of strongly correlated electron systems, and may play an important
role in the heavy fermion metals.~\cite{heavyfermionreview} This
interplay is not trivial even in the minimal system {\rm allowing it},
which consists of two localized spins ``imbedded''
into an electron Fermi sea.~\cite{Varmaetal}

\begin{figure}
 \epsfxsize=0.35\textwidth
% %\vspace*{0.3\textwidth}
 \epsfbox{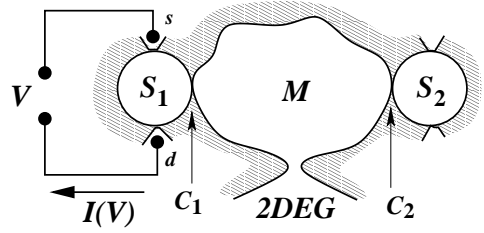}
\caption{The exchange interaction of $S=1/2$
  spins of dots $S_1$ and $S_2$ with the 2DEG results in
  the Kondo effect. The connection by weak contacts $C_1$ and $C_2$ to
  the bigger open dot $M$ creates the exchange (RKKY) interaction
  between the two spins. The current $I$ as a function of bias %voltage
  $V$ is measured between source ($s$) and drain ($d$) contacts.}
\label{fig:1}
\end{figure}

Until recently, such two-spin system was the subject of theoretical
investigations only. The quest for the practical implementations
of the quantum computing ideas has led to the interest in physics
of spin devices.
In this context, transport properties of two spin-carrying quantum
dots, coupled with each other by RKKY interaction, were studied
experimentally~\cite{Craig} in a device schematically shown in
Fig.~\ref{fig:1}. The authors of Ref.~\onlinecite{Craig}
were able to see the effect of RKKY interaction between the spins
by monitoring the Kondo-enhanced conductance through one of the dots.
Sufficiently strong RKKY interaction locks the localized spins into
a singlet or triplet state and destroys or weakens
the Kondo effect, thus liquidating the enhancement
of the conductance. The experiments\cite{Craig}
demonstrated the fact of spin coupling
and detailed quantitative measurements seem to be within the reach of
experimental capabilities. Further experiments will also bring
an exciting prospect of an experimental  investigation of
the interplay between RKKY interaction and Kondo effect
in a controllable setting.

In this Letter we develop a theory of conductance spectroscopy of the
spin states of two $s=1/2$ quantum dots coupled by the RKKY interaction.
We start with a characterization of the ground state and low-energy
excitation spectrum of the many-body system formed by the dots and
itinerant electrons of a two-dimensional electron gas (2DEG). Next we
concentrate on the case of a strong RKKY interaction, allowing us to
treat the Kondo effect perturbatively and elucidate the main features
of the $I$-$V$ characteristic of the device sketched in
Fig.~\ref{fig:1}. The measurement of differential conductance
$G(V)=dI/dV$
enables one to distinguish between the singlet and triplet ground
states of the two localized spins. We then consider the
crossover between the singlet and triplet states on the linear
conductance $G_0\equiv G(V=0)$. We identify signatures in the linear
conductance which may allow to distinguish the singlet-triplet
crossover from a quantum phase transition between these two states.
Finally, we investigate the sensitivity of the RKKY interaction to an
applied magnetic flux and determine the characteristic flux needed to
change the sign of the RKKY coupling.

The ground state of two localized spins interacting with
each other and with a Fermi-sea of itinerant electrons depends
sensitively on the relations between the corresponding interaction
constants. Under very special conditions,\cite{nfl}
requiring fine-tuning of the system parameters, a non-Fermi liquid
state of the system may be reached. Away from these special points in
the parameter space, the low-energy properties of the system are that
of a Fermi liquid, and we will concentrate on this generic case. The
ground state of the full system, including the localized and itinerant
spins, is a singlet. Variation of the exchange couplings leads to a
number of crossovers between different possible singlet states.  The
simplest Hamiltonian sufficient for describing the singlet states is
\begin{equation}
\begin{split}
& H =
J_1\hat {\bm s}_1 \hat{\bm S}_1 +
J_2\hat {\bm s}_2 \hat{\bm S}_2 +
J_{12}\hat {\bm S}_1 \hat{\bm S}_2.
\label{Hspin}
\end{split}
\end{equation}
Here ${\bm S}_1$ and ${\bm S}_2$ are the spins of the two dots
($S_{1,2}=1/2$), and ${\bm s}_\alpha =\sum_{\sigma\sigma'}
\psi^\dagger_{\alpha\sigma} {\bm \sigma}_{\sigma\sigma'}
\psi^{}_{\alpha\sigma'}$ are the local spin densities of itinerant
electrons. Electron operators $\psi_{\alpha\sigma}$ represent
electron states~\cite{GP} coupled to a localized spin in dot
$\alpha$.

First we note that at $J_{12}=0$ the Hamiltonian Eq.~(\ref{Hspin})
represents two independent $s=1/2$ Kondo systems characterized
by Kondo temperatures $T_{{\rm K}\alpha}\propto\exp(-1/\nu J_\alpha)$,
where $\nu$ is the band density of states at the Fermi level,
$\alpha=1,2$. At $T=0$, each of the two local spins is fully screened
by the corresponding modes of the itinerant electrons. In these
conditions, each of the dots provides a Kondo resonance for electron
tunneling.\cite{GP} The independent screening, giving rise to resonances
in tunnelling through each of the dots, remains in effect for a
sufficiently small inter-dot coupling. At stronger antiferromagnetic
coupling, $J_{12}\gg {\rm max}_\alpha \{T_{{\rm K}\alpha}\}$,
the two local spins form a singlet of its own, and
the Kondo resonances for the itinerant electrons vanish. In the case
of ferromagnetic ($J_{12}<0$) coupling, the two localized spins form a
triplet, which is fully screened by the itinerant electrons
interacting with the two dots and the Kondo resonances for
tunnelling through each of the dots persist.

\begin{figure}
\epsfxsize=0.40\textwidth
%\vspace*{0.3\textwidth}
\epsfbox{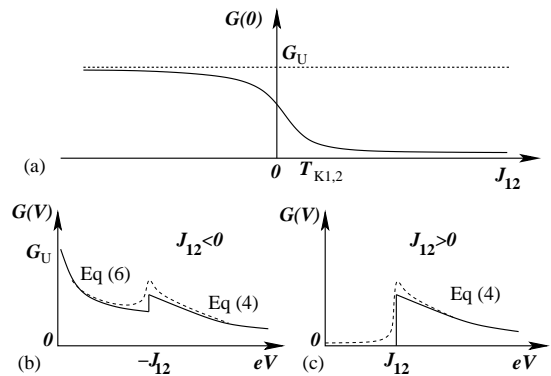}
\caption{(a) Linear conductance $G_0$ as a function of
  the RKKY coupling constant $J_{12}$ at $T=0$. The crossover from the
  triplet state $J_{12}<0$ at $|J_{12}|\gg {\rm max }_\alpha\{T_{{\rm
      K}\alpha}\}$ to a singlet state $J_{12}\gg {\rm
    max}_\alpha\{T_{{\rm K}\alpha}\}$ takes place at
  $|J_{12}|\lesssim{\rm max }_\alpha\{T_{{\rm K}\alpha}\}$ and results
  in the decrease of $G_0$ from a value close to the unitary limit
  $G_{\rm U}$ to a much smaller value controlled by the elastic
  co-tunnelling.  (b,c) Differential conductance $G(V)$ at $T=0$ and
  strong RKKY coupling, $|J_{12}|\gg {\rm max }_\alpha\{T_{{\rm
      K}\alpha}\}$. Solid lines: the second-order in exchange $J_1$
  result; dashed lines: schematic representation of the
  enhancement of $G(V)$ resulting from higher-order in $J_1$ calculation.}
\label{fig:2}
\end{figure}

The above consideration shows that the zero-temperature linear
conductance changes from a large value
(induced by the Kondo resonance)
at $(-J_{12})\gg {\rm max}_\alpha\{T_{{\rm K}\alpha}\}$, to a
small value at $J_{12}\gg {\rm max}_\alpha\{T_{{\rm K}\alpha}\}$.
In the over-simplified representation of the device sketched in
Fig.~\ref{fig:1} by the Hamiltonian (\ref{Hspin}), these asymptotes
are
\begin{equation}
G_{\rm U}=\frac{4e^2}{\pi\hbar}\frac{G_sG_d}{(G_s+G_d)^2}
\label{eq1}
\end{equation}
and $G=0$, respectively (here $G_s$ and $G_d$ are the conductances of
the junctions connecting the dot with the source and drain leads,
respectively).
In the special case~\cite{VH} of $T_{K2}=0$, the transition between
the two asymptotes is a jump at $J_{12}=0$ between
two Fermi liquid states. If $T_{\rm K2}$ is finite,
the transition shifts
to $J_{12}\sim T_{\rm K1}/\ln(T_{\rm K1}/T_{\rm K2})$
for $T_{\rm K2}\ll T_{\rm K1}$,
and remains at positive $J_{12}\lesssim {\rm max}_\alpha\{T_{{\rm K}\alpha}\}$.
This transition occurs via passing
through a non-Fermi liquid state,  which belongs to the same universality
class as the two-channel $s=1/2$ Kondo problem\cite{AL}. The
existence of such non-Fermi liquid state hinges on a special
particle-hole symmetry\cite{nfl} of the Hamiltonian
(\ref{Hspin}). However, in the case of generic parameters of a quantum dot
the Hamiltonian of the system also includes other terms ({\it
  e.g.} potential scattering terms leading to the elastic
co-tunnelling) which violate the required symmetry. In this case the
quantum phase transition between the two Fermi liquids is replaced
by a smooth crossover.\cite{PH,F,simon2005} The zero-temperature
conductance varies smoothly and monotonically with $J_{12}$, as
shown by the solid line in Fig.~\ref{fig:2}(a).\cite{cotunnel}

In the absence of the Zeeman splitting,
the energy of exchange interaction $J_{12}$ sets the threshold for the
inelastic electron scattering, accompanied by a change of the spin
state of two dots. Far above the
threshold, $|eV|\gg |J_{12}|$, the processes with flip of the spin
${\bm S}_1$ are allowed and, in the leading order, the differential
conductance $G(V)$ coincides
with the conductance of a single-quantum-dot device:
\begin{equation}
G(V)=\frac{3}{4} \frac{G_{\rm U}}{\ln ^2|eV/T_{{\rm K}1}|} .
\label{largebias}
\end{equation}
Here factor $3/4$ corresponds to the square of the operator of spin
$S=1/2$, and the logarithmic term represents the Kondo renormalization
of the exchange interaction in the weak coupling limit, $eV\gg
T_{K1}$.

Conductance Eq.~(\ref{largebias}) arises from the scattering amplitude
evaluated to the lowest order perturbation theory in the renormalized
interaction constant.
Within this accuracy, conductance has a sharp
step at $|eV|=|J_{12}|$. The height of the step depends on the sign of
$J_{12}$, {\it i.e.}, on wether the coupling of the two localized
spins is ferromagnetic  or antiferromagnetic.
In the case of ferromagnetic coupling, $J_{12}<0$,
conductance $G(V)$ is reduced by factor $3/2$ from the value given
by Eq.~(\ref{largebias}) if the bias is lowered just below the
threshold $|eV|=-J_{12}$. The renormalization group analysis
of the Hamiltonian (\ref{Hspin}) shows that after
the reduction at $|eV|=-J_{12}$, the differential
conductance again monotonically increases
as bias $V$ decreases:
\begin{equation}
G(V)=\frac{2G_{\rm U}}{\ln^2 |eV/T_{{\rm K}t}|}
\label{Glowbias}
\end{equation}
Here factor $2$ corresponds to the square of $S=1$ spin
operator and the logarithmic increase of $G(V)$
reflects the $S=1$ Kondo effect with the triplet Kondo temperature
$T_{{\rm K}t}$. In case $T_{{\rm K2}}\ll T_{{\rm K1}} \ll (-J_{12})$
an evaluation of the first logarithmic correction to the conductance
gives $T_{{\rm K}t}=T_{\rm K1}^2/|J_{12}|$, {\it i.e.}
at large $|J_{12}|$ the triplet Kondo temperature is much lower
than ${\rm max}_\alpha\{T_{{\rm K}\alpha}\}$.
For  the antiferromagnetic coupling between the two localized spins
the particle-hole symmetric Hamiltonian (\ref{Hspin}) yields zero
conductance.\cite{cotunnel}

\begin{figure}
 \epsfxsize=0.45\textwidth
%\vspace*{0.3\textwidth}
 \epsfbox{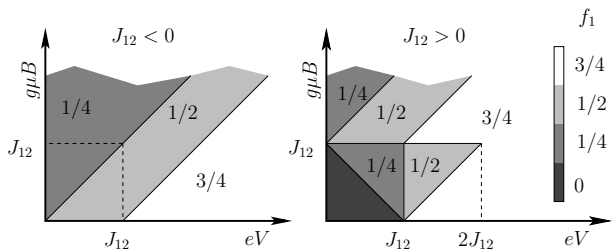}
\caption{The contour plot of $f_1(J_{12},V,B)$ is shown in the
bias ($V$) -- magnetic field ($B$) plane.
Function $f_1$ determines the differential conductance $G(V)$
at strong RKKY coupling $|J_{12}|\gg {\rm max }_\alpha\{T_{{\rm
K}\alpha}\}$, see Eq.~(\ref{GinMagField}).
}
\label{fig:3}
\end{figure}

Magnetic field may shift
and split the step in conductance $G(V)$.
The proper generalization of Eq.~(\ref{largebias}) to
include the Zeeman splitting $g\mu B$ by magnetic field $B$
reads
\begin{equation}
G(V)=\frac{f_1(J_{12},V,B)G_{\rm U}}
{\ln^{2}
[{\rm max}\{g\mu B, eV \}/T_{\rm K1}]},
\label{GinMagField}
\end{equation}
with a step-like function $f_1$, presented in Fig.~\ref{fig:3}.
At $g\mu B\gg |J_{12}|, |eV|$, the
localized spins are aligned along the field and $f_1=1/4$.
If $J_{12}<0$, Eq. (\ref{GinMagField}) must be supplemented
by condition $\max\{g\mu B,eV\}\gtrsim  \sqrt[3]{T_{{\rm K}1}^2
J_{12}}$; otherwise a $S=1$ Kondo effect develops,
see Eq.~(\ref{Glowbias}).

To assess the applicability of the above results obtained in the Born
approximation for the electron amplitude of tunnelling through the
dot, one may evaluate the next-order correction to that amplitude.
This correction diverges logarithmically at $|eV|=|J_{12}|$, similar
to the divergence occurring in the scattering off a single localized
spin in the presence of Zeeman splitting.\cite{Appelbaum} The
divergence is cut off by either the
temperature~\cite{Appelbaum} or the relaxation rate of the
localized spins in the out-of-equilibrium conditions.~\cite{Wolfle} As
the result, steps in the differential conductance, obtained within the
Born approximation, become asymmetric maxima of a finite width. Here
we present an estimate for the correction to Eq.~(\ref{Glowbias}) only
for the case of strongly asymmetric setup, {\it e.g.}, $G_s/G_d\ll 1$,
and assuming the lowest temperature, which allows for
the sharpest steps in the differential conductance. In these
conditions, the spin relaxation rate\cite{Korringa} responsible for
the cut-off of the divergency is of the order of $ 1/\tau_s\sim
|J_{12}|/[\ln^2(|J_{12}|/T_{{\rm K}1})].  $ The widths of the features
replacing steps in the differential conductance are of the order of
$\sim 1/\tau_s$, and the amplitudes of the corrections to steps
described by Eq.~(\ref{GinMagField}) are
\begin{equation}
\delta G\sim \frac{G_{\rm U}}{\ln^3(|J_{12}|/T_{{\rm
      K}1})}\ln\left[\ln^2\frac{|J_{12}|}{T_{{\rm K}1}}\right].
\label{correction}
\end{equation}
The amplitude of the correction is small if
$|J_{12}|\gg T_{{\rm K}1}$,
and the main effect of the higher-order terms of the perturbation
theory is the smearing of the steps. In order to determine the
nature of the ground state (singlet or triplet) by
conductance measurements in the presence of magnetic field, the
Zeeman splitting energy must exceed $1/\tau_s$ of
the features in the differential conductance.  The overall bias
dependence of the differential conductance at $B=0$ is sketched in
Fig.~\ref{fig:2}(b,c).

Linear conductance $G_0$ of the dot $S_1$ is determined
by the scattering $T$-matrix $T_{1}(\e)$ for $\e\lesssim T$:
\begin{equation}
G_0(T,J_{12})= G_{\rm U}
\int \frac{{\rm Im}T_{1}(\e) d\e}{4T\cosh^2(\e/2T)}.
\label{G.2}
\end{equation}
In the unitary limit $T_1(0)=i$. For $\e\sim T\gtrsim {\rm max}_\alpha\{T_{{\rm K}\alpha}\}$ we may use
the Born approximation~\cite{VGL}:
\begin{equation}
{\rm Im}T_1(\e)= \nu^2\cJ^2_\e \int
K(\e-\e')\frac{1+e^{\e/T}}{1+e^{\e'/T}}
d\e',
\label{delta2}
\end{equation}
where
%\begin{equation}
$
K(\omega) = 2\pi \sum_{\xi\xi'}\rho_\xi
\left|
\langle \xi' | {\bm S}_1 | \xi\rangle
\right|^2
\delta(\omega+E_\xi-E_{\xi'})
$
%\label{Komega}
%\end{equation}
is calculated with respect to the exact quantum
states $|\xi\rangle$ of the system of two localized spins,
$E_\xi$ is the energy of state $|\xi\rangle$,
and $\rho_\xi$ is the density matrix $\rho_\xi\propto \exp
(-E_\xi/T)$. The exchange constant ${\cal J}_\e$
is logarithmically renormalized by the Kondo effect:
${\cal J}_\e=1/[\nu \ln E/T_{{\rm K}1}]$ with
$E={\rm  max}\{\varepsilon,T,|J_{12}|\}$.

Within Born approximation, the linear conductance depends on $J_{12}$
only through the ratio $J_{12}/T$,
\begin{equation}
G_0 =
\frac{3}{2(3+e^{\cV/T})}
\left[1+\frac{\cV/T}{1-e^{-\cV/T}}
\right]
\frac{G_{\rm U}}{\ln^{2} (T/T_{{\rm K}1})}
.
\label{GKnoZeeman}
\end{equation}
The dependence of $G_0(T,J_{12})$ on $J_{12}$
has a maximum in the
region $|J_{12}|\lesssim T$ and conductance is higher on the triplet
side of the crossover.\cite{Tmax} (Note however, that the exponentially small
value of conductance far in the singlet region is an artefact of
our model.\cite{cotunnel})
At negative $J_{12}$, as  temperature $T$ decreases,
the conductance $G_0(T,J_{12})$ grows and
reaches $G_{\rm U}$ at $T\lesssim T_{\rm K1}^2/|J_{12}|$.
The shape of the $G_0(T,J_{12})$ vs. $J_{12}$
eventually approaches the step sketched in Fig.~\ref{fig:2}a. In the
generic case of the zero-temperature crossover between the two
fermi-liquid states, the width of the step saturates
and remains finite in the limit $T\to 0$.
This width is not universal and depends on the terms in the
exact Hamiltonian beyond the approximation of Eq.~(\ref{Hspin}).
If, however, the parameters of the system are tuned properly,
and the variation of $J_{12}$ takes the system through the non-Fermi-liquid
state at certain value of $J_{12}$, then the step width decreases
with lowering the temperature
as $\sqrt{T}$, see,
{\it e.g.}, [\onlinecite{delft}]. Finally, the limit $J_2=0$ within the model
Eq.~(\ref{Hspin}) corresponds to a sharp transition between two
Fermi-liquid states at $T=0$. A straightforward analysis, similar
to Ref.~[\onlinecite{VH}], yields the estimate $T_{{\rm K}1}/\ln (T_{{\rm K}1}/T)$
for the transition width at finite temperature.

The contact
interaction of the localized spins with the itinerant electrons of dot
$M$ at points $C_1$ and $C_2$, see Fig.~\ref{fig:1}, results in the
indirect exchange interaction between the localized spins. Unlike the
textbook RKKY interaction facilitated by freely propagating
electrons~\cite{Kittel}, here magnitude and sign of $J_{12}$ are
random, reflecting the chaotic electron motion in the dot $M$.
Roughly, the typical value of $J_{12}$ is
$\sim \mls J_{C_1} J_{C_2}$,
and the typical magnetic flux $\Phi_c$ needed for
changing $J_{12}$ substantially is $\sim\Phi_0$.
Here ${J}_{C_{1,2}}$ are the dimensionless constants of
the contact exchange interaction at
points $C_{1,2}$; $\mls=(\nu A)^{-1}$ is the mean level spacing
of one-electron energy levels in dot $M$,
$\nu$ is density of states of the 2DEG, $A$ is the
area of the dot $M$, and $\Phi_0=hc/e$.

The RKKY coupling $J_{12}$ may be expressed in terms of the scattering
matrix ${\cal S}_{1;2}$ of an electron propagating from contact $C_1$
to contact $C_2$:
\begin{equation}
\begin{split}
J_{12} =&  -2 J_{C_1}J_{C_2}\!\!\! \int \frac{d\e}{\pi} n(\e)
%\\
%&\times
{\rm Im}
\left\{
{\cal S}_{2;1}(\e)
{\cal S}_{1;2}(\e)
\right\}.
\end{split}
\label{JRKKY}
\end{equation}
Here $n(\e)$ is the Fermi function and
we assumed that electron propagation in $M$ is spin independent:
${\cal S}_{1\sigma;2\sigma'}(\e)=
\delta_{\sigma\sigma'}{\cal S}_{1;2}(\e)$.
Within the random matrix
theory, it is related in a standard way\cite{BRMP} to the one-electron
Hamiltonian $\hat {\cal H}$ of dot $M$. The ensemble average
$\langle J_{12}\rangle =0$, and we
calculate the correlation function~\cite{ZS} of the RKKY
constant $J_{12}(\Phi)$ over realizations of matrix $\hat {\cal
  H}(\Phi)$ at two values of magnetic flux $\Phi_{1,2}$ threading dot $M$,
\begin{equation}
\begin{split}
\langle J_{12}(\Phi_1)  J_{12}(\Phi_2) \rangle
 & =
\frac{\mls^2}{16 \pi^2}  {J}_{C_1}^2 {J}_{C_2}^2
\ln\left[
\frac{E_{\rm Th}^2}{{\cal E}_+{\cal E}_-}
\right].
\label{Jcorrf}
\end{split}
\end{equation}
Here
${\cal E}_\pm   =\gamma_{\rm esc}+
\kappa \mls (\Phi_1\pm \Phi_2)^2/\Phi_0^2
$; numerical factor $\kappa\sim 1$ depends on geometry;\cite{BRMP,Efetov}
$\gamma_{\rm esc}=N\mls/(2\pi)$ is the electron escape rate from
the middle dot into 2DEG through $N$ open channels;
$E_{\rm Th}= v_F/\sqrt{A}$ is the Thouless energy;
 and $v_{\rm F}$ is the Fermi
velocity. At $T\gtrsim {\cal E}_{\pm}$, ${\cal E}_\pm$ should be replaced by $T$ .

Equation (\ref{Jcorrf}) shows that the RKKY constant is symmetric with
respect to the inversion of magnetic flux $\Phi\to -\Phi$, and yields
the correlation flux value $ \Phi_{\rm c}^{J_{12}}\sim \Phi_0 $.
The flux $\Phi_{\rm c}^{J_{12}}$ is much larger than the correlation
flux $\Phi^{G}_{\rm c} \sim \sqrt{\gamma_{\rm esc}/E_{\rm Th}}
\Phi_0$ for the conductance of an open quantum dot\cite{BRMP}.
The difference between $\Phi_{\rm c}^{J_{12}}$ and
$\Phi_{\rm c}^{G}$ occurs
because the contribution to $\cV$ originates from energy levels
within the spectrum ``window'' $E_{\rm Th}$, whereas the conductance
of a dot is usually determined by levels within much
shorter energy interval $\gamma_{\rm esc}$.

In conclusion, the presented results may help one to determine
the spin states of quantum dots coupled by RKKY
interaction from transport measurements, see Eq.~(\ref{GinMagField})
and Fig.~\ref{fig:3}.
The evolution of the linear conductance with the variation of RKKY
interaction constant allows one to follow the transitions between
various ground states of the system. Finally, we found that the
magnetic field flux needed for variation of the RKKY coupling is of
the order of flux quantum $\Phi_0$.

Discussions with I.L. Aleiner, C.M. Marcus, and M.P.~Pustilnik are
gratefully acknowledged. The research was supported by
the MRSEC Program of the NSF under
award DMR 02-13282 and
by AFOSR grant F49620-01-1-0457 (MIT) and
by NSF grants DMR02-37296 and EIA02-10736
(University of Minnesota).


\begin{thebibliography}{99}

\bibitem{Craig}  N. J. Craig {\it et al.},
%J. M. Taylor, E. A. Lester, C. M. Marcus, M. P. Hanson, A. C. Gossard,
Science {\bf 304}, 565 (2004).

\bibitem{Hewson} A.C.~Hewson, {\it The Kondo
problem to Heavy Fermions} (Cambridge University Press, Cambridge, UK,
1997).

\bibitem{Kondoreview}
D. Goldhaber-Gordon {\it et al.}, Nature {\bf 391}, 156 (1998);
S.M.~Cronenwett {\it et al.},
%, T.H.~Oosterkamp and L.P.~Kouwenhoven,
Science {\bf 281}, 540 (1998);
J. Schmid {\it et al.}, Physica B {\bf 256-258}, 182 (1998).

\bibitem{Kittel}  C. Kittel, {\it Quantum Theory of Solids}
(Wiley \& Sons, New York, 1987), p. 360.

\bibitem{heavyfermionreview} A. Amato, Rev. Mod. Phys. {\bf 69},
1119 (1997).

\bibitem{Varmaetal}
C.~Jayaprakash {\it et al.},
%%%H.R.~Krishna-murthy, and J.W.~Wilkins,
Phys. Rev. Lett. {\bf 47}, 737 (1981);
B. A. Jones and C. M. Varma, Phys. Rev. Lett. {\bf 58},
843 (1987); %B. A. Jones and C. M. Varma,
Phys. Rev. B {\bf 40}, 324 (1989).

\bibitem{nfl}I. Affleck {\it et al.},
%A.W.W. Ludwig, and B. A. Jones,
Phys. Rev. B 52, 9528 (1995);
A. J. Millis {\it et al.},
%B. G. Kotliar, and B. A. Jones,
in {\it Field Theories in Condensed Matter Physics},
edited by Z. Tesanovic (Addison Wesley, Redwood City,
CA, 1990), p. 159.


\bibitem{GP} M. Pustilnik and L.I. Glazman,
J. Phys. Condens. Matter {\bf 16}, R513 (2004).

\bibitem{VH} M.~Vojta {\it et al.},
%R.~Bulla, and W.~Hofstetter,
Phys. Rev. B {\bf 65}, 140405(R) (2002).

\bibitem{AL} I. Affleck and A.W.W. Ludwig, Phys. Rev. B {\bf 48}, 7297
(1993).

\bibitem{PH} M.~Pustilnik {\it et al.},
%L.I.~Glazman, and W.~Hofstetter,
Phys. Rev. B {\bf 68}, 161303(R) (2003).

\bibitem{F} R.M. Fye, Phys. Rev. Lett. {\bf 72}, 916 (1994).

\bibitem{simon2005} The crossover in the system of Ref. [\onlinecite{Craig}] was studied numerically
by Pascal Simon, Rosa Lopez, and Yuval Oreg, Phys. Rev. Lett. {\bf
94}, 086602 (2005).

\bibitem{cotunnel}  The  co-tunnelling processes not accounted by
Hamiltonian~(\ref{Hspin}), result in a finite background conductance.

\bibitem{Appelbaum} J. Appelbaum, Phys. Rev. Lett. {\bf 17}, 91
(1966); J.A. Appelbaum, Phys. Rev. {\bf 154}, 633
(1967).

\bibitem{Wolfle} J. Paaske {\it et al.},
cond-mat/0401180.

\bibitem{Korringa}
This rate can be evaluated similar to the Korringa relaxation
rate, see L.I. Glazman and M. Pustilnik, cond-mat/0501007, p. 40.

\bibitem{VGL}
M.G. Vavilov {\it et al.},
%%%L.I. Glazman, and A.I.~Larkin,
Phys. Rev. B {\bf 68}, 075119 (2003).

\bibitem{Tmax}  Linear conductance $G_0(T,J_{12})$
at fixed $|J_{12}|\gg T_{{\rm K}1}$ has
a maximum at $T\simeq 0.3 |J_{12}| \sqrt{\ln{|J_{12}|/T_{{\rm K}1}}}$
due to competition between the Kondo effect and the RKKY interaction,
see {\it e.g.}, Ref.~[\onlinecite{VGL}].

\bibitem{delft} M. Pustilnik {\it et al.},
Phys. Rev. B {\bf 69}, 115316 (2004).

\bibitem{BRMP} C. W. J. Beenakker, Rev. Mod. Phys. {\bf 69}, 731
(1997); I.L.~Aleiner {\it et al.},
%P.W. Brouwer, and L.I. Glazman,
Phys. Reports {\bf 358}, 309 (2002).


\bibitem{ZS} Statistics of the RKKY
  coupling in metals
  was studied in A.Yu. Zyuzin and B.Z. Spivak, JETP
  {\bf 43}, 234 (1986).

\bibitem{Efetov} K. B. Efetov,  Phys. Rev. Lett. {\bf 74}, 2299 (1995).

\newpage

\end{thebibliography}
\end{document}